# An effective Nuclear Model: from Nuclear Matter to Finite Nuclei

**T.R.Routray[1]†, X.Vinas[2], S.K. Tripathy[3], M.Bhuyan[1], S.K.Patra[4], B.Behera[1]**
1. School of Physics, Sambalpur University, Jyotivihar, Sambalpur, Odisha-768019, INDIA.
2. Department d' Estructura I Constituents de la Materia, Universitat de Barcelona, Av. Diagonal 647, E-08028 Barcelona, Spain.
3. Depatment of Physics, Govt. College of Engineering Kalahandi, Bhawanipatna, Odisha-766002, INDIA.
4. Institute of Physics, Sachivalaya Marg, Bhubaneswar-751005, Odisha, INDIA.
E-mail: †trr1@rediffmail.com

**Abstract.** The momentum and density dependence of mean fields in symmetric and asymmetric nuclear matter are analysed using the simple density dependent finite range effective interaction containing a single Gaussian term alongwith the zero-range terms. Within the formalism developed, it is possible to reproduce the various diverging predictions on the momentum and density dependence of isovector part of the mean field in asymmetric matter. The finite nucleus calculation is formulated for the simple Gaussian interaction in the framework of quasilocal density functional theory. The prediction of energies and charge radii of the interaction for the spherical nuclei compares well with the results of other effective theories.

## 1. Introduction

The density dependence of nuclear symmetry energy $E_s(\rho)$ is a major area of contemporary nuclear research for its implications in understanding the nuclear equation of state (EOS) under extreme conditions of density, isospin asymmetry and temperature. However, another important dimension of nuclear EOS study is the momentum dependence of nuclear mean field. These two important aspects of nuclear EOS are interlinked and should be analysed simultaneously. Attempts in this direction have been made in the recent past [1, 2] in the analysis of flow data but not much success were achieved that may be attributed to the particulars of the experimental difficulties in the phenomena chosen and/or the deficiency in the analysis procedure. The nuclear mean field being the crucial quantity in the analysis procedure should have the capability of producing the desired density dependence of nuclear symmetry energy while the momentum dependent aspect is not changed and vice-versa. In this context we formulate a mean field in the frame work of non-relativistic mean field theory from a phenomenological finite range nucleon-nucleon effective interaction that has the aforesaid feature. Then we shall constrain the momentum dependence in Pure Neutron Matter (PNM) to a narrow region from thermal evolution study of nuclear matter properties. Any effective interaction applied in the nuclear matter (NM) regime need to be tested in the finite nucleus region for its acceptability and therefore we shall extend our study to the calculation of finite nucleus properties. In this connection we shall calculate energies and charge radii of 161-spherical nuclei.

## 2. Formalism

For the four general effective interactions direct $v_d^l(r), v_d^{ul}(r)$ and exchange $v_{ex}^l(r)$, $v_{ex}^{ul}(r)$ acting between pairs of like (l) and unlike (ul) nucleons which are functions of separation distance between the pair of interacting nucleons and depending on the total density of the medium, the energy density in NM at temperature $T$ can be expressed as,

$$H_T(\rho_n,\rho_p) = \frac{\hbar^2}{2M}\int [f_T^n(\vec{k}) + f_T^p(\vec{k})]\ k^2\ d^3k + \frac{1}{2}(\rho_n^2 + \rho_p^2)\int v_d^l(r)\ d^3r + \rho_n\rho_p\int v_d^{ul}(r)\ d^3r$$

$$+ \frac{1}{2}\iint [f_T^n(\vec{k}) f_T^n(\vec{k}') + f_T^p(\vec{k}) f_T^p(\vec{k}')]\ g_{ex}^l(|\vec{k}-\vec{k}'|)\ d^3k\ d^3k'$$

$$+ \frac{1}{2}\iint [f_T^n(\vec{k}) f_T^p(\vec{k}') + f_T^p(\vec{k}) f_T^n(\vec{k}')]\ g_{ex}^{ul}(|\vec{k}-\vec{k}'|)\ d^3k\ d^3k',$$

(1)

where, $g_{ex}^{l,ul}(|\vec{k}-\vec{k}'|)$ are Fourier transforms of the respective exchange interactions $v_{ex}^{l,ul}(r)$,

$$g_{ex}^{l,ul}(|\vec{k}-\vec{k}'|) = \int e^{i(\vec{k}-\vec{k}')\cdot\vec{r}}\ v_{ex}^{l,ul}(r)\ d^3r \qquad (2)$$

and the Fermi-Dirac momentum distribution function is given by

$$f_T^i(\vec{k}) = \frac{\xi}{(2\pi)^3}\frac{1}{1+\exp[\{\varepsilon_T^i(\vec{k},\rho_n,\rho_p)-\mu_T^i\}/T]} \qquad (3)$$

with $i = n, p$, where, $\xi = 2$ is the spin degeneracy factor, $\varepsilon_T^i(\vec{k},\rho_n,\rho_p)$ is the single particle energy, $\mu_T^i$ is the chemical potential of the nucleon and $\vec{k}$ is the momentum of the nucleon. The normalization factor $\frac{\xi}{(2\pi)^3}$ has been subject to the condition that integration over the momentum space of proton and neutron distribution functions shall result into respective densities $\rho_p$ and $\rho_n$,

$$\rho_i = \int f_T^i(\vec{k})\ d^3k,\ i = n, p. \qquad (4)$$

The finite range effective interaction used in this work is given by,

$$v_{eff}(\vec{r}) = t_0(1+x_0 P_\sigma)\delta(\vec{r}) + \frac{1}{6}t_3(1+x_3 P_\sigma)\left[\frac{\rho(\vec{R})}{1+b\rho(\vec{R})}\right]^\gamma \delta(\vec{r})$$

$$+ (W + BP_\sigma - HP_\tau - MP_\sigma P_\tau)f(r)$$

(5)

where, $f(r)$ is the functional form of a short range interaction of conventional form such as Yukawa, Gaussian or exponential and is specified by a single parameter $\alpha$, the range of the interaction. The density dependent term has been modified with the denominator containing the parameter $b$ in order to avoid supraluminous behaviour in NM. The remaining symbols in eq.(5) have their usual meaning. The interaction has altogether 11 number of parameters namely $t_0, x_0, t_3, x_3, b, \gamma, W, B, H, M$ and $\alpha$. The energy density in (1), for the interaction in (5) becomes,

$$H_T(\rho_n,\rho_p) = \frac{\hbar^2}{2M}\int \left[f_T^n(\vec{k}) + f_T^p(\vec{k})\right]k^2 d^3k + V_T(\rho_n,\rho_p) \qquad (6)$$

where, $V_T(\rho_n,\rho_p)$ is the interaction part given by

$$V_T(\rho_n,\rho_p) = \frac{1}{2}\left[\frac{\varepsilon_0^l}{\rho_0} + \frac{\varepsilon_\gamma^l}{\rho_0^{\gamma+1}}\left(\frac{\rho}{1+b\rho}\right)^\gamma\right](\rho_n^2 + \rho_p^2) + \left[\frac{\varepsilon_0^{ul}}{\rho_0} + \frac{\varepsilon_\gamma^{ul}}{\rho_0^{\gamma+1}}\left(\frac{\rho}{1+b\rho}\right)^\gamma\right]\rho_n\rho_p$$

$$+ \frac{\varepsilon_{ex}^l}{2\rho_0}\iint[f_T^n(\vec{k})f_T^n(\vec{k}') + f_T^p(\vec{k})f_T^p(\vec{k}')]g_{ex}(|\vec{k}-\vec{k}'|)d^3k\,d^3k'$$

$$+ \frac{\varepsilon_{ex}^{ul}}{2\rho_0}\iint[f_T^n(\vec{k})f_T^p(\vec{k}') + f_T^p(\vec{k})f_T^n(\vec{k}')]g_{ex}(|\vec{k}-\vec{k}'|)d^3k\,d^3k'.$$

(7)

The neutron, proton single particle potential can be obtained as the respective functional derivative of $V_T(\rho_n,\rho_p)$, i.e. $u_T^{n,p}(\vec{k}) = \frac{\partial V_T(\rho_n,\rho_p)}{\partial[f_\tau]}$, $\tau = n,p$ and is given by

$$u_T^{n(p)}(\rho_n,\rho_p,k) = \left[\frac{\varepsilon_0^l}{\rho_0} + \frac{\varepsilon_\gamma^l}{\rho_0^{\gamma+1}}\left(\frac{\rho}{1+b\rho}\right)^\gamma\right]\rho_{n(p)} + \left[\frac{\varepsilon_0^{ul}}{\rho_0} + \frac{\varepsilon_\gamma^{ul}}{\rho_0^{\gamma+1}}\left(\frac{\rho}{1+b\rho}\right)^\gamma\right]\rho_{p(n)}$$

$$+ \frac{\varepsilon_{ex}^l}{2\rho_0}\int f_T^{n(p)}(\vec{k}')\,g_{ex}(|\vec{k}-\vec{k}'|)\,d^3k' + \frac{\varepsilon_{ex}^{ul}}{2\rho_0}\int f_T^{p(n)}(\vec{k}')\,g_{ex}(|\vec{k}-\vec{k}'|)\,d^3k' + u_R(\rho)$$

(8)

where, $u_R(\rho)$ is the rearrangement energy that arises from the density dependence of the interaction and is given by

$$u_R(\rho) = \left[\frac{\varepsilon_\gamma^l}{\rho_0^{\gamma+1}}\frac{(\rho_n^2+\rho_p^2)}{2} + \frac{\varepsilon_\gamma^{ul}}{\rho_0^{\gamma+1}}\rho_n\rho_p\right]\frac{\gamma\rho^{\gamma-1}}{(1+b\rho)^{\gamma+1}}.$$

(9)

For complete study of asymmetric nuclear matter (ANM) we require to know all the nine parameters $\alpha, \gamma, b, \varepsilon_0^l, \varepsilon_0^{ul}, \varepsilon_\gamma^l, \varepsilon_\gamma^{ul}, \varepsilon_{ex}^l, \varepsilon_{ex}^{ul}$ where the new parameters are related to the interaction parameters as

$$\varepsilon_0^l = \rho_0\left[\frac{t_0}{2}(1-x_0) + 4\pi\alpha^3\left(W + \frac{B}{2} - H - \frac{M}{2}\right)\right], \tag{10a}$$

$$\varepsilon_0^{ul} = \rho_0\left[\frac{t_0}{2}(2+x_0) + 4\pi\alpha^3\left(W + \frac{B}{2}\right)\right], \tag{10b}$$

$$\varepsilon_\gamma^l = \rho_0^{\gamma+1}\left[\frac{t_3}{12}(1-x_3)\right], \tag{10c}$$

$$\varepsilon_\gamma^{ul} = \rho_0^{\gamma+1}\left[\frac{t_3}{12}(2+x_3)\right], \tag{10d}$$

$$\varepsilon_{ex}^l = 4\pi\alpha^3\rho_0\left(M - \frac{W}{2} + \frac{H}{2} - B\right), \tag{10e}$$

$$\varepsilon_{ex}^{ul} = 4\pi\alpha^3\rho_0\left(M + \frac{H}{2}\right). \tag{10f}$$

2.1. *Adjustment of interaction parameters*

It is necessary that the interaction parameters be fixed so as to give a good description of momentum dependence of the mean fields as well as the EOSs in SNM and PNM at zero-temperature. The energy densities of SNM and PNM at zero-temperature for the interaction in (5) with a Gaussian form of the finite range form factor $f(r)$ are,

$$H_0(\rho, T=0) = \frac{3\hbar^2 k_f^2 \rho}{10M} + \frac{(\varepsilon_0^l + \varepsilon_0^{ul})\rho^2}{4\rho_0} + \frac{(\varepsilon_\gamma^l + \varepsilon_\gamma^{ul})}{4\rho_0^{\gamma+1}}\left(\frac{\rho}{1+b\rho}\right)^\gamma \rho^2$$
$$+ \frac{(\varepsilon_{ex}^l + \varepsilon_{ex}^{ul})\rho^2}{4\rho_0}\left[\left(\frac{3\Lambda^6}{16k_f^6} - \frac{9\Lambda^4}{8k_f^4}\right) + \left(\frac{3\Lambda^4}{8k_f^4} - \frac{3\Lambda^6}{16k_f^6}\right)e^{-4k_f^2/\Lambda^2} + \frac{3\Lambda^3}{2k_f^3}\int_0^{2k_f/\Lambda} e^{-t^2} dt\right]$$
(11)

and

$$H_n(\rho, T=0) = \frac{3\hbar^2 k_n^2 \rho_n}{10M} + \frac{\varepsilon_0^l}{2}\frac{\rho_n^2}{\rho_0} + \frac{\varepsilon_\gamma^l}{2\rho_0^{\gamma+1}}\left(\frac{\rho}{1+b\rho}\right)^\gamma \rho_n^2$$
$$+ \frac{\varepsilon_{ex}^l \rho_n^2}{2\rho_0}\left[\left(\frac{3\Lambda^6}{32k_n^6} - \frac{9\Lambda^4}{8k_n^4}\right) + \left(\frac{3\Lambda^4}{8k_n^4} - \frac{3\Lambda^6}{16k_n^6}\right)e^{-4k_n^2/\Lambda^2} + \frac{3\Lambda^3}{k_n^3}\int_0^{2k_f/\Lambda} e^{-t^2} dt\right],$$
(12)

respectively, where $\Lambda = \frac{1}{\alpha}$. The Fermi momenta $k_f$ and $k_n$ are related to the density $\rho$ as $k_f^3 = 3\pi^2 \rho/2$ and $k_n^3 = 3\pi^2 \rho$. Out of the total nine parameters $\alpha, b, \gamma, \varepsilon_0^l, \varepsilon_0^{ul}, \varepsilon_\gamma^l, \varepsilon_\gamma^{ul}, \varepsilon_{ex}^l$ and $\varepsilon_{ex}^{ul}$ only six, namely $\alpha, b, \gamma, (\varepsilon_0^l + \varepsilon_0^{ul}), (\varepsilon_\gamma^l + \varepsilon_\gamma^{ul})$ and $(\varepsilon_{ex}^l + \varepsilon_{ex}^{ul})$ are required to describe the EOS of SNM. The parameters $\alpha$ and $(\varepsilon_{ex}^l + \varepsilon_{ex}^{ul})$ obtained through an optimization procedure so as to provide a correct momentum dependence of the mean field in SNM at normal density $\rho_0$ as demanded by optical model fits to nucleon-nucleus scattering data at intermediate energies [3-11] are $\alpha = 0.7582\,fm$ and $(\varepsilon_{ex}^l + \varepsilon_{ex}^{ul}) = -95.648$ MeV. In obtaining these parameters, we have used standard values of $Mc^2 = 939$ MeV, energy per nucleon at normal density in SNM $e(\rho_0) = -16.0$ MeV and $T_{f0} = 36.8\,MeV$ (corresponding to $\rho_0 = 0.16\,fm^{-3}$). The parameter $b$ is adjusted to avoid supraluminous behaviour of zero-temperature SNM at high densities [12]. The remaining two strength parameters $(\varepsilon_0^l + \varepsilon_0^{ul})$ and $(\varepsilon_\gamma^l + \varepsilon_\gamma^{ul})$ are determined from the saturation conditions. The exponent $\gamma$ determines the stiffness of the EOS of SNM at high densities. In figure 1(a) the pressure-density relations are shown for different values of $\gamma$ namely, $\gamma = 1/2, 1/3$ and $1/6$ all of which pass through the extracted band from the analysis of flow data in heavy-ion collision [13]. We use the value of $\gamma = 1/3$ that corresponds to a value of nuclear matter incompressibility, $K(\rho_0) = 227$ MeV at saturation in our the present study. The values of the six parameters in SNM are given in table 1. The effective nucleon mass, $\frac{M^*(k=k_f, \rho_0)}{M}$ in SNM at normal density, is predicted to be 0.71. As shown in figure 1(b), the momentum dependence of the mean field in SNM, $u_{ex}(k, \rho) = [u(k, \rho) - u(k_f, \rho)]$, obtained in the present case is in good agreement over a wide range of density with the microscopic results of UV14+UVII set of Wiringa [14].

**Table 1.** Values of interaction parameters in SNM.

| Parameters in SNM | $\gamma$ | $b\,[fm^3]$ | $\alpha\,[fm]$ | $(\varepsilon_{ex}^l + \varepsilon_{ex}^{ul})$ [MeV] | $(\varepsilon_\gamma^l + \varepsilon_\gamma^{ul})$ [MeV] | $(\varepsilon_0^l + \varepsilon_0^{ul})$ [MeV] |
|---|---|---|---|---|---|---|
| | 1/3 | 0.4184 | 0.758 | -95.648 | 110.744 | -112.749 |

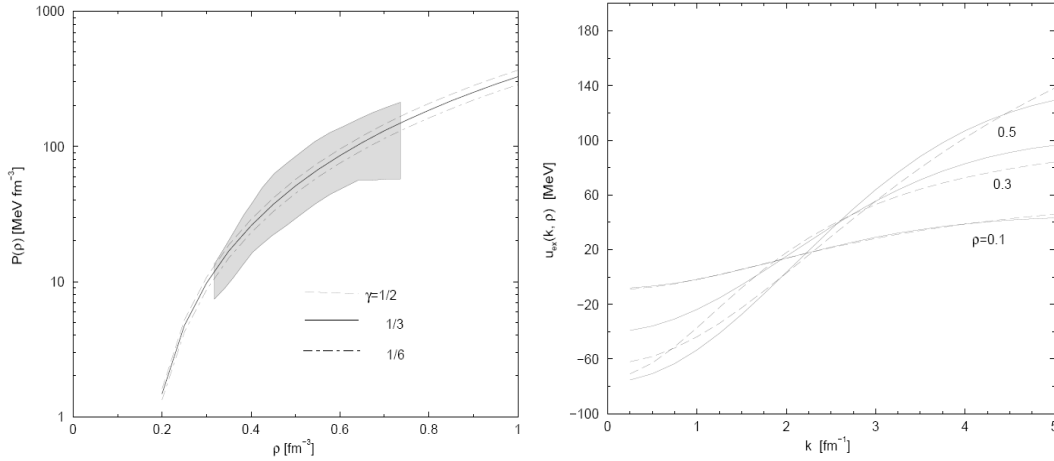

**Figures 1(a) &(b).** (a) Pressure-density relations for different values of $\gamma$. The experimentally extracted relation from the analysis of flow data from heavy-ion collision experiments is shown as the shaded region. (b) The momentum dependence of the mean field in SNM is compared with that of the microscopic calculations of Wiringa [14].

The complete calculation of EOS of PNM, requires the correct splittings of the three parameters $\left(\varepsilon_0^l + \varepsilon_0^{ul}\right)$, $\left(\varepsilon_\gamma^l + \varepsilon_\gamma^{ul}\right)$ and $\left(\varepsilon_{ex}^l + \varepsilon_{ex}^{ul}\right)$ into specific channels. However, there are no experimental or empirical constraints available on the splittings of these combined parameters except for the value of nuclear symmetry energy $E_s(\rho_0)$ at normal density. Different choices of these splittings can therefore lead to different density dependence of $E_s(\rho)$. The splitting of $\left(\varepsilon_{ex}^l + \varepsilon_{ex}^{ul}\right)$ into like (l) and unlike (ul) channels solely determines the neutron-proton effective mass splitting $(m_n^* - m_p^*)$ in ANM. Although, controversy exits on the nature of $(m_n^* - m_p^*)$ splitting amongst different models, it is largely accepted at present that the neutron effective mass in neutron rich matter will go over the proton effective mass. However, the magnitude of splitting still remains as an open problem. For our case in order to have the behaviour of $(m_n^* - m_p^*) > 0$, our splitting of $\left(\varepsilon_{ex}^l + \varepsilon_{ex}^{ul}\right)$ into like channel is restricted to the broad range $0 < \varepsilon_{ex}^l \leq \varepsilon_{ex}$, where, $\varepsilon_{ex} = \left(\varepsilon_{ex}^l + \varepsilon_{ex}^{ul}\right)/2$. In order to search for further constraining this broader range we have examined the thermal evolution of NM properties. The temperature dependence of the mean fields and the interaction parts of energy densities are simulated through the respective Fermi-Dirac momentum distribution functions while the interaction itself has no explicit temperature dependence. The momentum dependent parts of the mean fields involve the respective distribution functions and therefore imply self-consistent calculations. The distribution functions in SNM and PNM are evaluated at a given temperature $T$ and density $\rho$. The entropy density in SNM and for different values of $\varepsilon_{ex}^l$ within the range $0 < \varepsilon_{ex}^l \leq \varepsilon_{ex}$ in PNM have been calculated as a function of density at different temperatures. It is observed that the result obtained in the present case of Gaussian form is same as obtained in case of Yukawa form of the interaction given in Ref.[15]. The finding is that the entropy density in PNM exceeds that of SNM at a higher density if the parameter $\varepsilon_{ex}^l$ lies in between 0 to $2\varepsilon_{ex}/3$ and vice-versa is the case for $\varepsilon_{ex}^l$ lying between $2\varepsilon_{ex}/3$ and $\varepsilon_{ex}$. On the basis of the behaviour of the entropy density in PNM, the allowed range of $\varepsilon_{ex}^l$ is now divided into two groups, namely, (a) 0 to $2\varepsilon_{ex}/3$ and (b) $2\varepsilon_{ex}/3$ to $\varepsilon_{ex}$. For the specific choice of $\varepsilon_{ex}^l = 2\varepsilon_{ex}/3$ the entropy density in PNM approaches that of SNM asymptotically at large density. PNM being a system of one kind of particle, whether the entropy density in can exceed that of SNM which is a two component system or not. An answer to this basic question can further constrain the magnitude of neutron and proton effective mass splitting in ANM. In absence of a satisfactory answer to this question we proceed with the critical value $\varepsilon_{ex}^l = 2\varepsilon_{ex}/3$ in our subsequent calculations. The calculation of $m_n^*$ and $m_p^*$ in ANM require only the momentum dependent part of the mean fields which are shown as functions of asymmetry $\beta = (1 - 2Y_p)$ at the normal density $\rho_0$ and is shown in figure 2(a). For comparison we have also shown the DBHF[16], EBHF[17] and extended BHF [18] results. For the given splitting of $\left(\varepsilon_{ex}^l + \varepsilon_{ex}^{ul}\right)$, the splittings of the other two combinations $\left(\varepsilon_\gamma^l + \varepsilon_\gamma^{ul}\right)$ and $\left(\varepsilon_0^l + \varepsilon_0^{ul}\right)$ for interactions between two like and unlike nucleons can be described in terms of zero temperature nuclear symmetry energy $E_S(\rho_0, T = 0)$ and $E_S'(\rho_0, T = 0) = \rho \left. \frac{dE_s(\rho, T = 0)}{d\rho} \right|_{\rho = \rho_0}$ at normal nuclear matter density. Different theoretical models give similar values of $E_S(\rho_0, T = 0)$, but they differ widely in the values of $E_S'(\rho_0, T = 0)$ and therefore predict quite different high density behaviour of nuclear symmetry energy $E_S(\rho, T = 0)$ and equilibrium proton fraction $Y_P(\rho, T = 0)$ in beta-equilibrated $n + p + e + \mu$ matter (NSM). However, under beta-equilibrium the range of the functional $S^{NSM}(\rho, Y_p, T = 0) = \left[ (1 - 2Y_p(\rho, T = 0))^2 H_S(\rho, T = 0) \right]_{NSM}$ obtained from different theoretical models shows a much smaller variation over a wide range of density than exhibited by the respective symmetry energies $E_S(\rho, T = 0)$ and equilibrium proton fractions $Y_P(\rho, T = 0)$. In fact the functional $S^{NSM}(\rho, Y_p, T = 0)$ has a stiffest behaviour over a wide range of density which is universal to a good approximation in the sense that this behaviour remains almost the same for nuclear symmetry energies $E_S(\rho, T=0)$ whose high density behaviour is neither very stiff nor soft [19, 20]. In the present work we make use of this universal high density behaviour of the functional $S^{NSM}(\rho, Y_p, T=0)$ to constrain the density dependence of nuclear symmetry energy $E_S(\rho, T=0)$ at zero temperature. For this purpose we take a standard value of $E_S(\rho_0, T=0) = 34\,MeV$ [21, 22] and then vary $E_s'(\rho_0, T=0)$ to find out a critical value which gives the stiffest behaviour of $S^{NSM}(\rho, Y_p, T=0)$ over a wide range of high density. In the present case we get the value of $E_s'(\rho_0, T=0) = 24.4\,MeV$ that corresponds to the $L$ value $L = 3E_s'(\rho_0, T=0) = 73.2\,MeV$ in well agreement with the FRDM prediction [23]. The strength parameters thus obtained for PNM are given in Table-2. With the splittings of the three combined strength parameters $(\varepsilon_{ex}^l + \varepsilon_{ex}^{ul})$, $(\varepsilon_\gamma^l + \varepsilon_\gamma^{ul})$ and $(\varepsilon_0^l + \varepsilon_0^{ul})$ for interaction between two like and unlike nucleons, all the nine parameters necessary for a complete knowledge of ANM are fixed. The density dependence of symmetry energy is shown in figure 2(b) and is compared with the microscopic predictions. The effective interaction in (5) has eleven adjustable parameters and a study of neutron and proton mean fields and EOS of ANM can fix only nine parameters. The remaining two parameters are still open and can be fixed from finite nuclei.

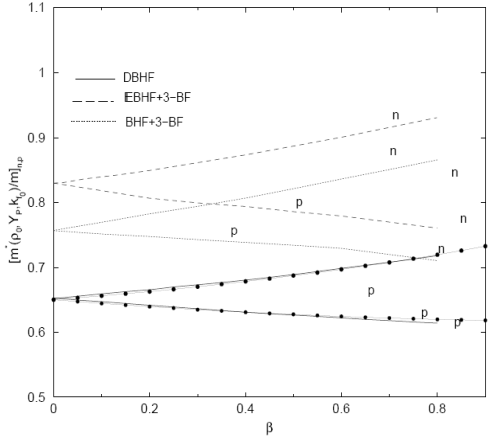 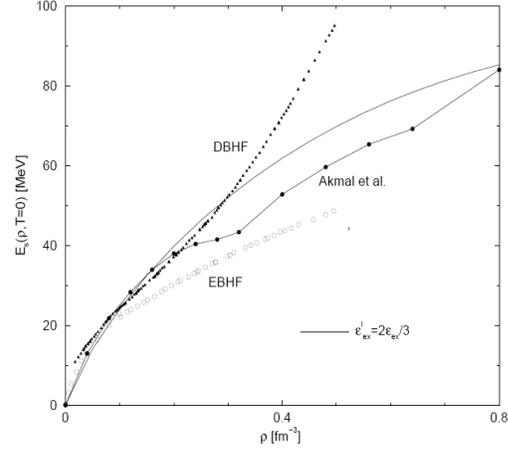

**Figure 2(a).** $m_n^*$ and $m_p^*$ as functions of neutron-proton asymmetry $\beta$. The curves with solid dots represent our present calculation normalised to the DBHF results at SNM.

**Figure 2(b).** Density dependence of Nuclear Symmetry Energy of the case $\varepsilon_{ex}^l = 2\varepsilon_{ex}/3$ and compared with the microscopic DBHF[16], EBHF[17] and BHF[18] calculations.

**Table 2.** Values of interaction parameters in PNM.

| $E_s'(\rho_0, T=0)$ | $\varepsilon_{ex}^l$ [MeV] | $\varepsilon_\gamma^l$ [MeV] | $\varepsilon_0^l$ [MeV] |
|---|---|---|---|
| 24.4 | -63.765 | 82.832 | -74.051 |

## 3. Finite Nuclei

The finite nucleus calculations with the effective interaction given in eq (5) with the Gaussian form shall be done in the frame work of quasilocal density functional theory (DFT), discussed in Refs. [24, 25]. Following these references, the quasilocal energy density functional is given by

$$E[\rho^{QL}] = \int H\, d^3r \tag{13}$$

where, $\rho^{QL}$ stands for the set $\{\rho_n, \rho_p, \tau_n, \tau_p, \vec{J}_n, \vec{J}_p\}$ and $\tau_q$ and $\vec{J}_q$ ($q = n, p$) are the (uncorrelated) kinetic energy and spin density build up with the auxiliary A-particle Slater determinant $\Psi_0$, where A is the mass number, which maps onto the local particle density $\rho(\vec{r}) = \sum_{i=1}^{A} |\varphi_i(\vec{r})|^2$. Therefore,

$$\tau_q(\vec{r}) = \sum_{i=1}^{A} \sum_\sigma \left|\vec{\nabla}\varphi_i(\vec{r}, \sigma, q)\right|^2 \tag{14}$$

$$\vec{J}_q(\vec{r}) = i \sum_{i=1}^{A} \sum_{\sigma,\sigma'} \varphi_i^*(\vec{r}, \vec{\sigma}', q)\, [(\vec{\sigma})_{\sigma,\sigma'} \times \vec{\nabla}]\varphi_i(\vec{r}, \vec{\sigma}, q). \tag{15}$$

The energy density $H$ in (13) is given by

$$H = \frac{\hbar^2}{2M}(\tau_n + \tau_p) + H_d^{nucl} + H_{ex}^{nucl} + H_d^{coul} + H_{ex}^{coul} + H^{SO} + H_{RC} . \quad (16)$$

The direct nuclear energy is given by

$$H_d^{nucl}(\vec{r}) = \frac{1}{2}\int d^3r' \left[ (W + \frac{B}{2})\rho(\vec{r})\rho(\vec{r}') - (H + \frac{M}{2})[\rho_n(\vec{r})\rho_p(\vec{r}') + \rho_p(\vec{r})\rho_n(\vec{r}')] v_d^{finite}(|\vec{r} - \vec{r}'|) \right] \quad (17)$$

In the exchange integral, the extended Thomas-Fermi approximation upto $\hbar^2$ is used that results into the exchange energy density,

$$H_{ex}^{nucl} = H_{ex,0}^{nucl} + H_{ex,2}^{nucl} \quad (18)$$

where, the first term corresponds to the zeroth order approximation,

$$H_{ex,0}^{nucl} = \int d^3s \, v_{ex}^{finite}(s) \left[ \frac{1}{2}\left(M + \frac{H}{2} - B - \frac{W}{2}\right) \sum_{q=n,p} \left(\rho_q(r) \frac{3j_1(k_q s)^2}{k_q s}\right) + \left(M + \frac{H}{2}\right)\rho_n(r) \frac{3j_1(k_n s)}{k_n s} \rho_p(r) \frac{3j_1(k_p s)}{k_p s} \right]$$

$$(19)$$

with $\vec{s} = \vec{r} - \vec{r}'$ being the relative co-ordinate and $k_q(\vec{r}) = [3\pi^2 \rho_q(\vec{r})]^{1/3}$ and $j_1$ is the 1st order spherical Bessel function. The second term corresponds to $\hbar^2$ correction

$$H_{ex,2}^{nucl}(\vec{r}) = \sum_q \frac{\hbar^2}{2M} \left\{ (f_q - 1)\left(\tau_q - \frac{3}{5}k_q^2 \rho_q - \frac{1}{4}\nabla^2 \rho_q\right) + k_q f'_q \left(\frac{1}{27}\frac{(\nabla \rho_q)^2}{\rho_q} - \frac{1}{36}\nabla^2 \rho_q\right) \right\}, \quad (20)$$

with $f_q = f_q(\vec{r}, k_q)$ and $f'_q = \left(\frac{\partial f_q(\vec{r},q)}{\partial k}\right)_{k=k_q}$, $f_q(\vec{r}, k_q)$ being the inverse of the position and momentum dependent effective mass given by

$$f_q(\vec{r}, k) = 1 + \frac{M}{\hbar^2 k}\frac{\partial V_{ex,q}^{nucl}(\vec{r}, k)}{\partial k} \quad (21)$$

where $V_{ex,q}^{nucl}$ is Wigner transform of the exchange potential

$$V_{ex,p}^{nucl} = \int d^3s \, e^{i\vec{k}\cdot\vec{s}} \left[ \left(M + \frac{H}{2} - B - \frac{W}{2}\right)\rho_p(r)\frac{3j_1(k_p s)}{k_p s} + \left(M + \frac{H}{2}\right)\rho_n(\vec{r})\frac{3j_1(k_n s)}{k_n s} \right]. \quad (22)$$

Coulomb energy is taken as usual as the direct plus exchange contribution and spin orbit is computed using a zero range force similar to that used in Skyrme and Gogny forces. The last part of our energy density considered comes from the zero-range part of the interaction,

$$H_{RC} = \frac{t_0}{4}\left[(1-x_\sigma)\left[\rho_n^2 + \rho_p^2\right] + (4+2x_0)\rho_n\rho_p\right] + \frac{t_3}{24}\left[(1-x_3)(\rho_n^2+\rho_p^2) + (4+2x_3)\rho_n\rho_p\left(\frac{\rho}{1+b\rho}\right)^\sigma\right] \quad (23)$$

Applying now the variational principle to the functional in (13) with the single-particle orbitals , the single particle equations are obtained to be

$$h_q \varphi_i = \varepsilon_i \varphi_i \quad (24)$$

where,
$$h_q = -\vec{\nabla}\frac{\hbar^2}{2m_q^*(r)}\vec{\nabla} + u_q(\vec{r}) - i\vec{W}_q(\vec{r})(\nabla \times \sigma) \quad (25)$$

with
$$\frac{\hbar^2}{2m_q^*(\vec{r})} = \frac{\delta E}{\delta \tau_q(\vec{r})}, \quad u_q = \frac{\delta E}{\delta \rho_q(\vec{r})} \quad \text{and} \quad \vec{W}_q(\vec{r}) = \frac{\delta E}{\delta \vec{J}_q}. \quad (26)$$

The single particle states are the solutions of (24) for the $E$ in (13) calculated for the energy density $H$ in (16). In calculating $E$ there are still two open parameters of the interaction apart from the spin-orbit strength parameter $W_0$. Here we have considered $t_0$ and $x_0$ as the open interaction parameters which are determined from the binding energy of $^{40}$Ca and $^{208}$Pb and $W_0$ from the splitting of the $1p$ level in $^{16}$O. In this procedure adopted to fix the two open interaction parameters, the nuclear matter predictions do not change. With all the interaction parameters determined, we have calculated the binding energy of 161 even-even spherical nuclei between $^{16}$Ne and $^{224}$U. The pairing correlations within the BCS approach using a density dependent delta force is taken into account as well as the two-body centre of mass correction. In figures 3(a) and (b), we plot the differences in the calculated and experimental energies and charge radii, respectively, as function of mass number A. The root mean square deviation in energy, rmsE, and the corresponding deviation in radius, rmsR, are 2.609 and 0.0168 respectively. The rmsE value is comparable to Gogny D1S [26], BCP2[27] and NL3[28]. The rmsR compares with that of Sly4[29] and BCP1[27]. The energy distribution, however, shows a slope having underbound nature in lighter and medium-heavy mass region that need to be examined with the variation of NM parameters with in their standard uncertainties.

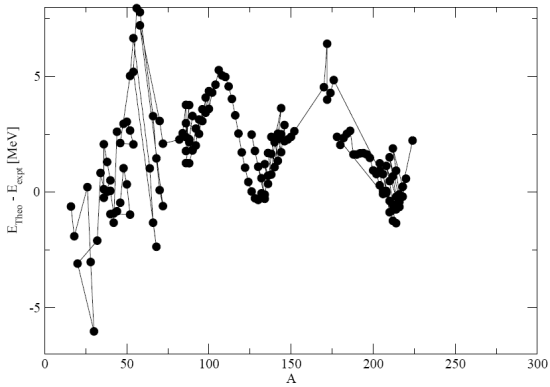
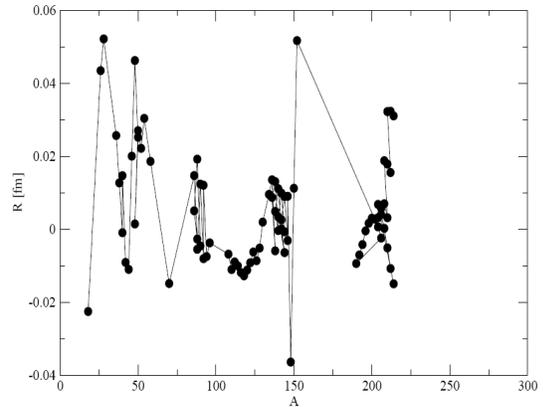

**Figure 3(a).** Differences between the theoretical and experimental energies.

**Figure 3(b).** Difference between the theoretical and experimental charge rms radii.

**Acknowledgments**

This work is supported by the grant No UGC-DAE-CSR/KC/2009/NP-06/1354 dated 31-07-2009 and is covered under the SAP programme of School of Physics, Sambalpur University.